\documentclass[12pt,twoside,showpacs,preprintnumbers,amsmath,amssymb]{revtex4}

\font\tenit=cmti10
\usepackage{graphicx}
\usepackage{dcolumn}
\usepackage{bm}

\def\dx{\displaystyle}
\def\epsilon{\varepsilon}

\def\div{\hbox{\rm div}\, }

\newcommand{\bigint}{\tenit\int}

\def\section#1{{\par\vglue 0.6cm
\noindent {\bf#1}\vglue 0.2cm\par}}
\def\subsection#1{\par\vglue 0.2cm
{\noindent{\it#1}\vglue 0.1cm}}

\begin{document}

\title{ Is the supersolid superfluid?}

\author{Dimitar I. Pushkarov}
 \altaffiliation{  e-mail dipushk@issp.bas.bg}
\affiliation{ Institute of Solid-State Physics,
 Bulgarian Academy of Sciences, Sofia 1784, Bulgaria and\\ Fatih University, Istanbul 3400}
\begin{abstract}
An analysis of previous theories of superfluidity of quantum solids
is presented in relation to the nonclassical rotational moment of inertia (NCRM)
found first in Kim and Chan experiments. A theory of supersolidity
is proposed based on the presence of an additional conservation law.
It is shown that
the additional entropy or mass fluxes depend on the quasiparticle
dispersion relation and vanish in the effective mass approximation.
This implies that at low temperatures when the parabolic part of the
dispersion relation predominates the supersolid properties should be
less expressed.
\end{abstract}

\maketitle

\section{Introduction}
The experiments of Kim and Chan \cite{KC1} breathed new life into the old idea of possible superfluidity of solids. A quantum solid possessing superfluid properties has been called a supersolid. Originally a supersolid should be a crystalline body where a nondissipative mass current can occur. This should correspond to the superfluid state of liquid helium (helium-II) observed by Kapitsa, and explained theoretically first by Landau. The superfluidity is now well studied and a number of effects has been found, predicted and explained. Between them is the change of the rotational moment due to the fact that the superfluid fraction cannot be involved into rotation at velocities less than the critical one. The qualitative explanation of such a behavior according to Landau is that at small velocities no excitations can be generated. A successful hydrodynamical description is given by the so-called two-fluid (or two-velocity) hydrodynamics. From a mathematical point of view the new element in the two-velocity hydrodynamic equations is the potentiality of the superfluid velocity $\mathbf{v}_s$ which reads $ \mathbf{v}_s = \nabla \mu$ with $\mu$ for the chemical potential in the frame where $\mathbf{v}_s=0$. As a result, a new vibrational mode, the second sound, appears. The phase transition into a superfluid state is well defined and the corresponding changes of the thermodynamic characteristics are well investigated.

The quantum-mechanical consideration connects the superfluidity with the Bose-Einstein condensation (BEC). Later on, such kind of condensation in the momentum space was observed in some gases as well. This is the reason to talk on a macroscopic quantum state described by the condensate wave function.

In their works Kim and Chan have observed a nonclassical rotational moment (NCRM), i.e. a rotational moment of inertia which changes its value with temperature in a way bearing a resemble with the Kapitza experiments
with rotating liquid helium. They argue that this is enough to conclude that the body has been in a supersolid state and that the superfluidity has been finally observed in all three states of matter (gas, liquid and solid). Lately, the term supersolid has become a synonym of a body with NCRM.

The first reasonable question is whether the NCRM implies superfluidity (supersolidity). Is the supersolid state "superfluid" or this is an evidence of a new phenomena, maybe more interesting and famous than superfluidity, but nevertheless, of different kind.
The existing experimental observations and theoretical analysis did not give an unambiguous answer yet.

Originally, the concept of supersolid appeared for a crystalline body inside which a nondissipative (macroscopic) mass current can exist. First considerations (Penrose and Onsager, Andreev and Lifshits, Legget, Chester etc.) had a crystalline bodies in mind. Defects in such crystals are imperfections of \emph{the crystal lattice}, or lattice with an ideal periodicity but with a number of atoms less than the number of the lattice sites (Andreev-Lifshits). The first question is therefore if the experiments can be understood from  such a point of view. Most probably this is not the case.

Let us first consider the validity of the Landau derivation of the critical velocity. In liquid helium, the energy in the frame where the superfluid velocity is zero can be written in the form:
\begin{equation}\label{Landau}
    E = E_0 + \mathbf{P}_0 \mathbf{v}+ \frac{1}{2} Mv^2, \quad \mathbf{P} = \mathbf{P}_0 + M\mathbf{v}
\end{equation}
The same relation for an elementary excitation $\varepsilon(p)$ reads
\begin{equation}\label{Land2}
    E = \varepsilon(p) + \mathbf{p} \mathbf{v}+ \frac{1}{2} Mv^2
\end{equation}
where $\mathbf{p}$ is the momentum in the frame where $\mathbf{v}=0$.
The least possible change in energy due to the excitation created is $\varepsilon(p) -pv$ and should be negative in order to reduce the energy of the system. This yields
\begin{equation}\label{Land3}
    v > \varepsilon(p)/p .
\end{equation}
It is worth noting that equation (\ref{Landau}) is always valid because it follows directly from the Galilean principle for \emph{macroscopic} quantities. Relation (\ref{Land2}) corresponds to the \emph{microscopic} characteristics of the \emph{elementary excitation}. In a homogeneous and isotropic (Newtonian) space these two relations coincide. However this is not the case in a crystalline solid where quasiparticle states are classified with respect the quasimomentum, not momentum. Quasimomentum is simply a quantum number which apears due to the periodicity of the lattice. There are \emph{no Galilean transformations} for quasiparticle characteristics. The transformation relations which replace the Galilean ones were derived in \cite{AP85,Singapore,Nauka,PhysRep}). The macroscopic momentum (the mass flux) is not a mean value of the quasimomentum, but of the product $\dx m \frac{\partial \varepsilon}{\partial \mathbf{k}}$ with $\mathbf{k}$ for the quasimomentum. In addition, phonons in crystals have zero momentum and do not transfer mass in contrast to the phonons in liquids. All this implies that the Landau criterion (\ref{Land3}) does not work in crystalline bodies. Its 'quasiparticle' analogue should look like
\begin{equation}\label{}
    v^{-1} > \frac{m}{\varepsilon} \frac{\partial \varepsilon}{\partial \mathbf{k}} = m \frac{\partial \ln{\varepsilon}}{\partial \mathbf{k}}
\end{equation}
 or
\begin{equation}\label{}
   m v < \frac{\partial\ln{\varepsilon}}{\partial \mathbf{k}} .
\end{equation}
But, there is still a question what is, say, the phonon qusimomentum in the co-moving (with the superfluid fraction) frame. In addition, $m=0$ for acoustic phonons.
To avoid any misunderstanding, let us stress again that whatever the dispersion relation of the elementary excitations and the spectrum classification parameter (momentum or quasimomentum), the macroscopic fluxes have to obey Galilean relation~(\ref{Landau}).

Next, it is very important that the conservation laws (which are the basis of the hydrodynamics) exist only in an \emph{inertial laboratory frame}. And this laboratory frame is privileged, not Galilean (see more details in Ref.~\cite{PhysRep}).

If one considers Bose-condensation of \emph{quasiparticles} then such a condensate in a crystal must be characterized by a value of the \emph{quasimomentum} in a privileged frame.

Finally, the particles or quasiparticles (say vacancions) undergoing Bose-condensation should interact weakly enough. It was shown \cite{meStat} that the vacancion gas is most ideal near the middle of the energy band, not in the bottom.

It is seen, therefore, that the situation in a crystalline body is completely different compared to liquids and gases.

Nevertheless, the first hydrodynamical theory of the superfluidity of solids \cite{AL69} was developed in a close analogy with the Landau theory of helium II. Andreev and Lifshits introduced two velocities for a normal and a superfluid fraction of the solid and apply the potentiality condition for the superfluid velocity, $\mathbf{v}_s = \nabla \mu$. They used the Galilean invariance, so
in the frame, where the superfluid
component is in rest ($\mathbf{v}_s = 0$), the energy per unit volume is:
\begin{equation}\label{}
E = \rho {v_s}^2/2 + \mathbf{ p}.{\bf v}_s + \epsilon, \quad
\mathbf{ j} = \rho \mathbf{v}_s + {\bf p}
\end{equation}
where $\mathbf{j}$ is the momentum per unit volume equal to the mass flow while
$\mathbf{p}$ is the momentum in the frame with $\mathbf{v}_s =0$,
  \begin{equation}\label{}
\epsilon = \epsilon( S,\rho, w_{ik})
\end{equation}
is the internal energy as a function of the entropy, density and the distortion
(not symmetric) tensor
$$
  w_{ik} = \frac {\partial u_{i}}{\partial x_k}.
$$
   The tensor of small deformations is as usually equal to
$$ u_{ik} =\frac{1}{2}\left\{\frac {\partial u_{i}}{\partial x_k}+
   \frac {\partial u_k}{\partial x_l}\right\}
$$
and its trace equals the relative variation of the volume
$$
u_{ii} = w_{ii} = \delta V/V
$$
A new point is that this trace is now not connected to the density variation with the known relation, i.e.
\begin{equation}\label{}
w_{ii} \ne - \frac {\delta \rho}{\rho}
\end{equation}
 In this notation,
\begin{equation}\label{}
d\epsilon = TdS + \lambda_{ik}w_{ik} + \mu d\rho + ({\bf v}_n
 - {\bf v}_s) d{\bf p}.
\end{equation}

 A standard procedure follows based on the conservation laws:
\begin{equation}\label{}
 \dot \rho + \div {\bf j} = 0, \qquad
\frac {\partial j_i}{\partial t} + \frac {\partial \Pi_{ik}}
{\partial x_k} = 0.
\end{equation}
\begin{equation}\label{}
{\dot S} + \div (S{\bf v}_n + {{\bf q} / T}) = {R /  T}, \quad (R > 0 )
\end{equation}
\begin{equation}\label{}
  {\dot {\bf v}}_s + \nabla \varphi = 0.
\end{equation}
The unknown quantities $\Pi_{ik}, \varphi, {\bf q}, R $ have to be determined
so as to satisfy the redundant energy conservation law:
\begin{equation}\label{}
  \dot E + \div {\bf Q} = 0.
\end{equation}
  The time derivative of $E$ reads:
\begin{eqnarray}
\dot E &=& T \dot S +
\lambda_{ik}{\frac {\partial {\dot u}_{k}}{\partial
 x_k}} - \mu \div {\bf j} - \div \left({ \frac {{v_s}^2}
 {2}}{\bf j} \right) + {\bf j} \nabla {\frac {{v_s}^2}{2}}
 \\
 &-& ({\bf j} - \rho {\bf v}_n) \nabla \varphi - v_{ni}
	{\frac {\partial \Pi_{ik}}{\partial x_k}} +
  {\bf v}_n {\bf v}_s \div {\bf j}
 \nonumber\\
 &=& - \div \left( {\bf j}{{{v_s}^2} \over{ 2}} + ST{\bf v}_n +
  {\bf v}_n ({\bf v}_n {\bf p}) \right)
  + T(\dot S + \div S{\bf v}_n )
  \nonumber\\
  &+& \lambda_{ik} {\frac {\partial {\dot u}_k}  { \partial x_i}}
  + ({\bf j} - \rho {{\bf v}_{n}})  \nabla  \left( \varphi -
{{{v_s}^2} \over 2} \right) - \rho {\bf v}_n \nabla \mu
	\nonumber\\
     &-& v_{ni} {\frac {\partial} {\partial x_k}}
 \left\{
 \Pi_{ik} -
  \rho v_{si}v_{sk} + {v_{si} p_k} + v_{sk} p_i
  \right.
  \nonumber\\
    &+& \left.
   [-\epsilon +TS + ({\bf v}_n - {\bf v}_s){\bf p} + \mu \rho]
  \delta_{ik} \right\} - \mu \div {\bf j}.
  \nonumber
  \end{eqnarray}
  \par
  Here, a term of the form
 $
 \dx
  v_{ni} \lambda_{kl} \frac {\partial w_{kl} }{\partial x_i}
 $
is neglected as cubic in ``normal motion''.
With the aid of conservation laws the time derivative of energy was written in the form \cite{AL69}:
  \begin{eqnarray} \label{Edot}
\dot E &+& \div \left\{ \left( {{{v_s}^2}\over 2} + \mu \right) {\bf j}
 + ST{\bf v}_n + {\bf v}_n ({\bf v}_n {\bf p}) + {\bf q} + \varphi ({\bf j}
  - \rho {\bf v}_n) + \right.
  \nonumber\\
   &+& \left.
    v_{nk}\pi_{ki} - \frac {{}_{}^{}}{} \lambda_{ik}{\dot u}_k \right\} =
  \nonumber\\
	&=& R + \pi_{ik} \frac {\partial v_{ni}}{\partial x_k} +
   \psi \div ( {\bf j} - \rho {\bf v}_n) + {{{\bf q} \nabla T} \over T}
   + ( v_{nk} - {\dot u}_k) \frac {\partial \lambda_{ik}}{\partial x_i},
	   \end{eqnarray}
\par
\noindent
   This yields	the following expressions for the fluxes:
\begin{eqnarray}\label{Pi-ik}
\Pi_{ik} &=&
\rho v_{si} v_{sk} + v_{si}p_k + v_{nk} p_i
    \nonumber\\
   &+& [- \epsilon + TS + ({\bf v}_n - {\bf v}_s) {\bf p} +
   \mu \rho ] \delta_{ik} - \lambda_{ik} + \pi_{ik},
    \\
  \varphi &=& {{v_s}^2 \over 2} + \mu + \psi.
    \end{eqnarray}
	\begin{eqnarray}\label{Q}
 {\bf Q} &=& \left( {\frac {{v_s}^2}{2}} + \mu \right)
 {\bf j} + ST{\bf v}_n +
  {\bf v}_n ({\bf v_n}{\bf p}) +  \mathbf{q}
   \nonumber\\
   &+& \psi({\bf j} - \rho {\bf v}_n) + v_{nk} \pi_{ki} -
\lambda_{ik} {\dot u}_k
      \end{eqnarray}
  and the dissipation function of the crystal is
\begin{eqnarray}
   R = - \pi_{ik}{\frac {\partial v_{ni}}{\partial x_k}} -
  \psi \div({\bf j}- \rho {\bf v}_n) -
     {{{\bf q}\nabla T}\over T} - (v_{nk} - {\dot u}_k)
  {\frac {\partial \lambda_{ik}}{\partial x_i}}
\end{eqnarray}
\par
\noindent We shall not write here the relations between
$\pi_{ik}, \psi, q$ and $(\mathbf{v_n - \dot u})$ that follow
from the Onsager principle and the positivity of the
dissipative function. The main consequence is that with neglecting dissipation one has
$\mathbf{v}_n = \mathbf{\dot  u}$. The normal motion is therefore the motion of
lattice sites (which may not coincide with given atoms). The superfluid flow could,
 hence, be possible at a given (even not moving) lattice structure.

However, instead of (\ref{Edot}) the time derivative $~\dot E$~ can also be
written in the form:
  \begin{eqnarray}\label{Edot2}
\dot E &+& \div \left\{ \left( {{{v_s}^2}\over 2} + \mu \right) {\bf j}
 + ST{\bf v}_n + {\bf v}_n ({\bf v}_n {\bf p}) + {\bf q} + \psi ({\bf j}
  - \rho {\bf v}_n) + \right.
    \\
   &+& \left.
v_{nk}\pi_{ki} - \frac {{}_{}^{}}{} \lambda_{ik}{v}_{nk} \right\} =
    \nonumber \\ &=& R + \pi_{ik}
\frac {\partial v_{nk}}{\partial x_i} +
\psi \div ( {\bf j} - \rho {\bf v}_n) + {{{\bf q} \nabla T} \over T}
   +  \lambda_{ik}\frac {\partial}{\partial x_i}({\dot u}_k - v_{nk}),
    \nonumber
\end{eqnarray}
\par\noindent
which leads to other expressions for fluxes.

In this case the  nondissipative theory yields:
  \begin{eqnarray}
\dot E &\!\!+\!\!& \div \left\{ \left( {{{v_s}^2}\over 2} + \mu \right) {\bf j}
 + ST{\bf v}_n + {\bf v}_n ({\bf v}_n {\bf p})	+
 (\varphi \! -\! \mu \! - \! \frac {v_{s}^{2}}{2})
({\bf j}
  - \rho {\bf v}_n) + \right.
 \nonumber \\  & \!\! + \!\! &
    v_{nk}
\left[	\frac{}{} \Pi_{ki} -
\rho v_{si} v_{sk} + v_{sk}p_i + v_{ni} p_k - \right.
    \\
   &\!\!- \!\!&
  \left. \left.
 [- \epsilon + TS + ({\bf v}_n - {\bf v}_s) {\bf p} +
   \mu \rho ] \delta_{ik} \frac{}{}
 \right]
  \frac {{}^{}}{{}_{}}\right\} =
    \nonumber \\ &\!\!=\!\!&
\left\{  \frac{}{} \Pi_{ki} -
\rho v_{si} v_{sk} + v_{sk}p_i + v_{ni} p_k - \right.
    \nonumber\\
   &\!\!- \!\!&
  \left. \left.
 [- \epsilon + TS + ({\bf v}_n - {\bf v}_s) {\bf p} +
   \mu \rho ] \delta_{ik}
  \frac {{}^{}}{{}_{}}
  \right\} \frac{\partial v_{nk}}{\partial x_i} \right.
   + \nonumber \\ &\!\! + \!\! &
 (\varphi - \mu - \frac {v_{s}^{2}}{2})
    \div ( {\bf j} - \rho {\bf v}_n) +
     \lambda_{ik}\frac {\partial {\dot u}_k}{\partial x_i}
    \nonumber
\end{eqnarray}
\par\noindent
and hence,
\begin{eqnarray}\label{Pi-ik2}
\Pi_{ik} &=&
\rho v_{si} v_{sk} + v_{si}p_k + v_{nk} p_i +
   \left[- \epsilon + TS + ({\bf v}_n - {\bf v}_s) {\bf p} +
   \mu \rho
  \frac {{}^{}}{{}}
 \right] \delta_{ik}
  \\
\varphi &=& \mu + \frac {v_{s}^{2}}{2}, \quad \qquad \lambda_{ik} = 0
 \quad {\hbox { !!!}}
   \end{eqnarray}
The procedure used is, therefore, not unique. The relation
$\mathbf{v}_n = \mathbf{\dot u}$ was not derived, but presupposed.
In fact, the consideration started as a three-velocity theory
($\mathbf{\dot u}, \mathbf{v_n}, \mathbf{v_s}$) and the identity
$\mathbf{\dot u} = \mathbf{v_n}$ follows from a condition the time
derivative of the total energy be not dependent on the
$\frac{\partial \lambda_{ik}}{\partial x_k}$ which is not well
grounded. Next, the conservation laws are written in the system where
$\mathbf{v_s} = 0$
and this is not the laboratory frame in which the lattice cites are in their equilibrium positions.


That is why we turned to another approach based on our theory of the quasiparticle
 kinetics and dynamics in deformable crystalline bodies
 \cite{PhysRep,Singapore,Nauka,Pushk2fluid}.
This theory works with exact (in the frame of the quasiparticle approach) selfconsistent
set of equations including the nonlinear elasticity theory equation
and a transport Boltzman-like equation valid in the whole Brillouin
zone of quasiparticles with arbitrary dispersion law. The theory is
developed for crystalline bodies
subject to time-varying deformations and arbitrary velocities.

\section {Partition Function and Thermodynamic Relations}

Let us consider a gas of quasiparticles with dispersion law
~$ \epsilon (\bf k)$~ at low temperatures, when the frequency of normal processes is much larger that of the Umklapp processes, i.e.
$$
\tau_{n}^{-1} \gg \tau_{U}^{-1} .
$$
The distribution function ~$ n_{k}({\bf k,\, r}, \,t)$~ corresponds
to ~$ S_{max}$~  with conserved energy $E$, quasiparticle density
$n$, quasimomentum $\mathbf{K}$ and momentum (mass flow) $\mathbf{j}$ defined, respectively, as:
\begin{equation}
S({\bf r},\, t) =\bigint
 \frac{{}^{}}{{}_{}} s[n_{k}]\, d {\bf k},
 \end{equation}
  where
 $
s[n_{k}]  = (1+ n_{k}) \ln (1 + n_{k}) - n_{k} \ln n_{k}
$
 \begin{eqnarray}
E({\bf r},\, t) & =& \int \limits_{}^{}
 \epsilon_{k} \, n_{k}\, d {\bf k}
 \\
n({\bf r},\, t) & =& \int  \limits_{}^{} n_{k}\, d {\bf k}
 \\
{\bf K}({\bf r},\, t) & =& \int  \limits_{}^{}{\bf k} \, n_{k} \, d {\bf k}
 \\
{\bf j}({\bf r},\, t) & =& m \int \limits_{}^{} \frac {\partial \epsilon_{k}}
{\partial {\bf k}} \,
n_{k} \, d {\bf k} ,\qquad  d{\bf k} = \frac {1}{(2 \pi)^3}
 \,d k_{1} d k_{2} d k_{3}
\end{eqnarray}
\par \noindent
This yields
$$
n_{k}(\mathbf{k, \, r}\, t) = \left\{
\exp \left( \frac {\epsilon_{k} - \mathbf{V.k} - m \mathbf{ W}.
 (\partial \epsilon_{k}/{\partial \mathbf{k}})
 - \mu}{T} \right) - 1 \right\}^{-1}
$$
with $\mathbf{V, W}$ and $\mu$ for Lagrangian multipliers.
Varying ~$S$~ yields
$$
T\delta S = \delta E - \mathbf{ V.}\delta \mathbf{ K} - \mathbf{ W.} \delta \mathbf{ j}
 - \mu \, \delta n
$$
$$
\Omega = E - TS - \mathbf{ V.K} - {\bf W .j } - \mu n
$$
and respectively
$$
d \Omega =  - S d T - \mathbf{K.}d \mathbf{V} - \mathbf{j .}d\mathbf{ W  } - n \, d \mu
$$
The nondissipative equations involve the following conservation laws:
\begin{eqnarray}
 {\dot n} &+& \div \mathbf{ J} = 0,
  \qquad \mathbf{J} = \mathbf{ j}/m
 \\
 \frac {\partial { j  }_{i}}{\partial t} &+&
  \frac {\partial \Pi_{ik}}{\partial x_k}
 = 0,
 \\
  \frac {\partial { K }_{i}}{\partial t} &\!+\!&
 \frac {\partial L_{ik}}{\partial x_k}
  = 0,
 \\
 {\dot S } &+& \div \mathbf{ F } = 0,
\nonumber \\
 {\dot E } &+& \div \mathbf{ Q } = 0
\end{eqnarray}
\par \noindent
To second order with respect to velocities ~$ \mathbf{ V}$~ and ~$\mathbf{  W}$~
one has:
\begin{equation}\label{}
J_{i} =  n^{0} V_{i} +  n_{ij}W_j,
\end{equation}
\par \noindent
where

\begin{equation}\label{}
 n_{ij} = \int n_{k}^{0} \, \nu_{il}
({\bf k}) \, d {\bf k}, \qquad
\nu_{il}(\mathbf{ k}) = m \frac {\partial^{2} \epsilon_{k}}{\partial k_i
\partial k_l}
\end{equation}

\par\noindent
The local drift velocity $ \mathbf{U}= \mathbf{j}/m n^0$ is then
\begin{equation}\label{}
U_i = V_i + \frac {n_{il}}{n^{0}} W_{l}
\end{equation}
The mass flux is, therefore, not collinear to any of velocities $\mathbf{V}$ and $\mathbf{W}$. Analogously,
\begin{equation}\label{}
K_i = \rho_{il}V_l + m n^0 W_i,
\end{equation}
\begin{equation}\label{}
 \rho_{il} = - \int k_i k_l \frac {\partial n_{k}}{\partial \epsilon_{k}}\,
d \mathbf{ k} =
  \left. \frac {\partial^2 \Omega}{\partial V_i \partial V_l}
  \right|_{T, \mu, \mathbf{ W}}
\qquad
 {\rho^{-1}}_{il} =
  \left. \frac {\partial^2 E}{\partial K_i \partial K_l}
  \right|_{S,n,\mathbf{ j}}
\end{equation}
To second order in velocities \textit{the diagonal terms of the quasimomentum flux
tensor coincide with the thermodynamic potential} $\Omega(T, \mathbf{V}, \mathbf{W}, \mu) $ \cite{Singapore,PhysRep,Nauka}:
\begin{equation}\label{}
 L_{ij} = \int k_i \frac {\partial \epsilon_{k}}{\partial k_{j}}
 n_k \, d{\bf k} = \Omega^0 \, \delta_{ij}
\end{equation}
and the momentum flux tensor has the form:
\begin{equation}\label{}
\Pi_{il} = - \Omega_{il} = T \int \ln( 1 + n_{k}^{0}) \, \nu_{il}({\bf k})\, d \mathbf{k}
\end{equation}
\noindent
The energy flux is:
\begin{equation}\label{}
Q_{i}({\bf r},t) = \bigint \epsilon_{k}
 \frac {\partial \epsilon_{k}}{\partial k_{i}} n_{k}\, d {\bf k}
 = W^{0} V_i +(TS_{il} + \mu \, n_{il})W_l
\end{equation}

where
\begin{equation}\label{}
S_{il} = \int s[n_{k}^{0}]\, \nu_{il} \, d{\bf k}
\end{equation}
and
$W^0 = E^0 -\Omega^0$ is the enthalpy at $\mathbf{V = W} = 0$.
Hence, $ W^{0} \mathbf{V}$ is the energy flux known from the classical hydrodynamics, and there are additional terms due to the supersolid behavior.
The full hydrodynamic system consists then of four equations:
\begin{equation}\label{H1}
{\dot n} + n \, \div \mathbf{ U} = 0,
\end{equation}
\begin{equation}\label{H2}
mn {\dot U}_i - \frac {\partial \Omega_{il}}{\partial x_l} = 0
\end{equation}
\begin{equation}\label{H3}
\rho_{is}
 \frac {\partial \Omega_{sl}}{\partial x_l}
- n \frac {\partial \Omega^{0}}{\partial x_{i}}
 + n^2 (\delta_{il} - \beta_{il}){\dot W}_{l}
= 0
\end{equation}
\begin{equation}\label{H4}
{\dot E} + W^{0} \div {\bf U} + TS \left(
 \frac {S_{il}}{S} - \frac {n_{il}}{n} \right)
 \frac {\partial W_{l}}{\partial x_{i}} = 0
\end{equation}
where
\begin{equation}\label{H5}
\beta_{il} = \rho_{ik} n_{kl} / (n^{0})^2
\end{equation}
Taking into account the thermodynamic identity:
\begin{equation}\label{H6}
d E =T\,dS + \mu \, d n + {\bf V.}d{\bf K} + {\bf W.} d{\bf j}
\end{equation}
\par\noindent
the energy conservation law can be replaced by entropy equation:
\begin{equation}\label{}
{\dot S} + S\,\div {\bf U} + S \left(
 \frac {S_{il}}{S} - \frac {n_{il}}{n} \right)
 \frac {\partial W_{l}}{\partial x_{i}} = 0
\end{equation}
It is seen that the mass flux and the entropy flux have different
velocities both in magnitude and direction. This means that a mass
flux without entropy transport can take place. This implies an existence of a superfluid density~$\rho^s$.
\medskip

\textbf{In the case of quadratic dispersion law}
\begin{equation}\label{}
\rho_{il} = m\, n^{0} \delta_{ik}, \quad
 \frac {S_{il}}{S} = \frac {n_{il}}{n}, \quad
 \beta_{il} = \delta_{il}, \quad j_i = \nu_{il}K_l
\end{equation}
and the additional entropy flux vanishes. This implies that the
superfluid effects should be negligible at very low temperatures where the excitations with parabolic dispersion relation predominate.

Let us now rewrite the hydrodynamic set in terms of Landau superfluid
 theory in order to see better the analogy.

\section {Cubic crystal}
Let us first consider, for simplicity, a cubic crystal. Then, to the
second order with respect to velocities one has
\begin{equation}\label{}
{\bf J}({\bf r},t) = n \,{\bf V} + \nu \, {\bf W}, \qquad
 {\bf Q}({\bf r},t) = W^{0}{\bf V} + (p+q){\bf W}
\end{equation}
\begin{equation}\label{}
\Pi_{ik} = p\,\delta_{ik}, \qquad
L_{ik} = - \Omega ^{0}\,\delta_{ik}, \qquad
{\bf F} = S {\bf V} + p_{T} {\bf W}
\end{equation}
where
\begin{eqnarray}
p ({\bf r}, t) &=& \frac {1}{3} m \int \left( \frac {\partial \epsilon_{k}}
{\partial {\bf k}} \right)^2  n_{k}^{0}\, d {\bf k},
\quad \rightarrow \quad \!\! -\frac {m}{m^*} \Omega^0
 \label{p}\\
\nu ({\bf r}, t) & =& \frac {1}{3} m  \int  \frac {\partial^{2} \epsilon_{k}}
 {\partial k^{2}}\, n_{k}^{0}\, d {\bf k},
  \qquad \rightarrow \quad \,\, \frac {m}{m^*}n
 \label{nu}\\
 q ({\bf r}, t) &=& \frac {1}{3} m  \int \epsilon_{k}
 \frac {\partial^{2} \epsilon_{k}}
 {\partial k^{2}}\, n_{k}^{0}\, d {\bf k},
  \quad \rightarrow \quad \,\, \frac {m}{m^*} E^0
 \label{q}
\end{eqnarray}
and the following relations take place:
\begin{equation}\label{}
p_{\mu} = \left( \frac {\partial p}{\partial \mu} \right)_{T} = \nu,
\qquad
p_{T} = \left( \frac {\partial p}{\partial T } \right)_{\mu} =
   \frac {p + q - \mu \nu}{T}
\end{equation}
The meaning of the quantities involved can be seen from their limiting expressions in the effective mass ($m^*$) approximation.

In the notation of Landau theory:	~$\dx \mathbf{ F } = S \mathbf{ V}^n, \,\,
\mathbf{ V} = \mathbf{V}^s$~
and the system of equations (\ref{H1}--\ref{H5}) takes the form:
 \begin{equation}\label{}
 {\dot n} + n^s \div {\bf V}^s + n^n \div \mathbf{V}^n = 0, \qquad
 {\dot S} + S \div \mathbf{ V}^n = 0,
 \end{equation}
\begin{equation}\label{}
 { \mathbf{\dot K}} + S \nabla T + n \nabla \mu = 0, \qquad
\frac {\partial \mathbf{ j}}{\partial  t} + p_{T} \nabla T +
  p_{\mu} \nabla \mu = 0
\end{equation}
where \quad $ \dx  n^n = S \frac {p_{\mu}}{p_{T}}, \quad n^s = n - n^n$~
and the number density flux is
\begin{equation}\label{}
\mathbf{J} = n^s \mathbf{ V}^s + n^n \mathbf{ V}^n
\end{equation}
\section {Second Sound}
\bigskip
\par
One has in variables ~$\mu, \, T, \, \mathbf{ V}^s, \, \mathbf{ V}^n$
\begin{eqnarray}
\alpha {\dot T} + \beta {\dot \mu} + n^s \nabla \mathbf{ .V}^s +
 n^n \mathbf{ \nabla .V^n} &=& 0
\nonumber\\
\gamma {\dot T} + \alpha {\dot \mu} + S \nabla .\mathbf{ V}^n &=& 0
\nonumber\\
S \nabla T + n \nabla \mu + \rho^s \mathbf{ \dot V}^s + \rho^n \mathbf{ \dot V}^n &=& 0
\nonumber\\
n S \nabla T + n n^n \nabla \mu = \rho^n n^s \mathbf{ \dot V}^s +
 \rho^n n^n \mathbf{ \dot V}^n &=& 0
\nonumber
\end{eqnarray}
where
\begin{equation}\label{}
\alpha = \left.\frac {\partial n}{\partial T}\right|_{\mu} =
\left.\frac {\partial S}{\partial \mu}\right|_{T} , \qquad
\beta =
 \left.\frac {\partial n}{\partial \mu }\right|_{T } , \qquad
\gamma = \left.\frac {\partial S}{\partial T}\right|_{\mu}
\end{equation}
\begin{equation}\label{}
\rho^n = m n S/p_T, \qquad \rho = \frac {1}{3} \int \mathbf{ k}^2 n^0
 (1 + n^0)\, d\mathbf{ k}, \qquad
  \rho^s = \rho - \rho^n
\end{equation}
\begin{equation}\label{}
\mathbf{ K} = \rho \mathbf{ V} + mn \mathbf{ W} = \rho^s \mathbf{ V}^s + \rho^n \mathbf{ V}^n
\end{equation}
For {\it quasiparticles with quadratic dispersion law} ~$ n^s = \rho^s = 0$:
\begin{eqnarray}
\alpha {\dot T} + \beta {\dot \mu} + n \nabla \mathbf{ .V}^{n} &=& 0
\\
\gamma {\dot T} + \alpha {\dot \mu} + S \nabla .\mathbf{ V}^n &=& 0
\\
S \nabla T + n \nabla \mu + \rho^s \mathbf{ \dot V}^s + \rho \mathbf{ \dot V}^n &=& 0
\end{eqnarray}
\par\noindent
If the number of quasiparticles is not conserved
 \begin{equation}\label{}
 \omega_{0}^2 (\mathbf{ q}) = \frac{TS}{C_{v} \rho} \mathbf{q}^2
 \end{equation}
If the number of quasiparticles is conserved
\begin{eqnarray}
\omega^2 (\mathbf{q}) & =& \omega^{2}_{0} (\mathbf{ q})
 \left\{ \left( 1 - \frac {\alpha n}{ \beta S} \right)^2 +
 \frac {C_v n^2}{T \beta S^2} \right\}
\nonumber\\ &=&
 \left[ \frac {T}{C_v} \left(\frac{\partial s}{\partial v_{0}}\right)_{T}^2
 - \left( \frac {\partial \mu}{\partial v_{0}}\right)_{T}
   \right] \frac {\mathbf{q}^{2}}{\rho}
  =
   \left( \frac {\partial P}{\partial n}\right)_{s} \frac {n}{\rho}\, \mathbf{q}^2
\end{eqnarray}
where
 \quad $ s = S/n, \quad v_0 = 1/n$~.


\section{Quasiparticles with non-quadratic dispersion law}
\bigskip
\par\noindent
If the number of quasiparticles is not conserved
\begin{equation}\label{}
\omega^{2}(\mathbf{ q}) = (1+ \delta^{\rho}) \frac {TS^2}{C_v \rho^{n}} \mathbf{q}^2
  ,\quad
  \delta ^{\rho} = \frac {\kappa ^{\rho} \kappa^n }{\kappa^{\rho} - \kappa^n},
 \quad \kappa^{\rho} = \frac {\rho^s}{\rho^n}, \quad
 \kappa^{n} = \frac {n^s}{n^n}
\end{equation}
If the number of quasiparticles is conserved
\begin{eqnarray}
\omega_{1}^2 (\mathbf{ q})& =& \delta^n \left\{ \omega_{0}^2 (\mathbf{ q},\,
 \rho \! =\!
 \rho^n) - \frac {1}{n \rho^{n}}\,
 \frac {T S}{C_{v}}\,
 \frac {(\partial P/\partial T)^{2}_{n}}
 {(\partial P/\partial n)_{s}}\, \mathbf{q}^2 \right\}
\\
\omega_{2}^2(\mathbf{q}) & =& (1 - \kappa^n) \, \omega^2(\mathbf{q}, \, \rho \! =\! \rho^n) +
  \delta^n  \omega_{0}^2 (\mathbf{q},\,\rho \! =\! \rho^n)
  - \omega_{1}^2 (\mathbf{q})
 \\  &-&
 2 \kappa^n \frac {T S}{\rho^{n} C_{v}}\,
  \left( \frac {\partial P}{\partial T}\right)_n \mathbf{q}^2
\nonumber
\end{eqnarray}
  \begin{equation}\label{}
  \omega_2(\mathbf{q}) > \omega_1(\mathbf{q})
  \end{equation}
\section{Conclusion}
It is shown that the theory of superfluidity of solids should not be a replica of the Landau theory of superfluidity. For crystalline bodies a two-velocity theory of supersolidity is presented with accounting of the quasimomentum conservation law. Such a theory cannot be applied to disordered systems, glasses etc.

\section{Acknowledgements}
The author thanks Professor V.Kravtsov for the inviation to the Abdus Salam ICTP, Trieste, where this work was submitted. Acknowledgements are also given regarding the partial financial support from the National Science Fund, Contract F-1517



\begin{references}
\bibitem{KC1}  
E. Kim and M.H.W. Chan, Nature, \textbf{427} 225 (2004);
E. Kim and M.H.W. Chan, Science \textbf{305} 1941-44 (2004);
E. Kim and M.H.W. Chan, J.Low Temp. Phys. \textbf{138} 859 (2005)
\bibitem{AP85}
A.F.Andrev and D.I.Pushkarov, - \emph{ Sov.Phys.JETP} \textbf{ 62} (5) 1087-1090 (1985)
\bibitem{Singapore}  
 D.I.Pushkarov, - Quasiparticle theory of defects in solids, World Scientific, Singapore, 1991
\bibitem{Nauka}   
 D.I.Pushkarov, Defektony v kristallakh (Defectons in crystals - quasiparticle
approach to quantum theory of defects) in Russian,  Nauka, Moscow  1993
\bibitem{PhysRep} 
 D.I. Pushkarov, - Phys.Rep. \textbf{354} 411 (2001)
\bibitem{meStat}
D.I.Pushkarov, - Phys.Stat.Sol.(b) \textbf{133} 525 (1986)
\bibitem{Pushk2fluid} 
D.I.Pushkarov and R. Atanasov, Phys.Scripta \textbf{42} 481 (1990)
\bibitem{AL69} 
A.F.Andreev and I.M.Lifshitz - Zh.Eksp.Teor.Fiz. \textbf{56} (12) 2057 (1969)
\end{references}
\end{document}